\documentclass[twocolumn,secnumarabic,amssymb, nobibnotes, aps, prd]{revtex4}
\usepackage{graphicx}

\begin{document}
\title{Comment on "Vortices induced in a superconducting loop by asymmetric kinetic inductance and their detection in transport measurements"}
\author{V.L. Gurtovoi  and  A.V. Nikulov}
\affiliation{Institute of Microelectronics Technology and High Purity Materials, Russian Academy of Sciences, 142432 Chernogolovka, Moscow District, RUSSIA.} 
\begin{abstract} The paper by G. R. Berdiyorov, M. V. Milosevic, and F. M. Peeters [Phys. Rev. B 81, 144511 (2010)] studies theoretically the dynamic properties of a superconducting loop. The authors claim that their consideration of asymmetric loop relates to our experimental results in this field. We point out that this claim is incorrect and explain shortly the true paradoxical essence of the results of our measurements of asymmetric superconducting loop. 
\end{abstract}

\maketitle

\narrowtext

\section{Introduction}
The authors of the paper \cite{Peeters} investigate theoretically the flux quantization and vortex entry in a thin rectangular superconducting loop (with thickness much smaller than the coherence length $d \ll \xi$ and the penetration depth $d \ll \lambda$, width $w = 2\xi $ and lateral sizes $a = 32\xi $ and $b = 16\xi $) under influence of a transport electric current applied through the normal contacts size equal to the loop width and of a magnetic field $B$ perpendicular to the plane of the loop. They refer in the Introduction on some results of transport measurements of asymmetric superconducting loop \cite{Physica1967}, including our one \cite{PM2001,JETP07Sh}, making urgent their theoretical investigation of the complexity of the problem rising if asymmetry is introduced.  Below they note about a similarity of behaviour of the output voltage on the applied field obtained in their paper \cite{Peeters} and experimentally in our work \cite{Letter07}. In the Conclusion the authors claim that they theoretical results \cite{Peeters} relate to our experimental efforts in the field \cite{JETP07Sh,Letter07}. We must say that this claim is a consequence of misunderstanding of our experimental results \cite{PM2001,JETP07Sh,Letter07} as well as the results of other authors \cite{Physica1967}. We should explain in this Comment that the method used in \cite{Peeters} is not valid for a description both the critical current oscillations \cite{JETP07Sh} and the dc voltage oscillation \cite{Physica1967,Letter07}. It can not used also for a description of the Little-Parks oscillations \cite{Letter07}. In order to avoid the muddle we use the term "Little-Parks oscillations" only for the resistance oscillations observed first by Little and Parks \cite{LP1962}.

\section{Transition from continuous to strong discrete spectrum of permitted states}
First of all we should note that a notion about vortex entry used in \cite{Peeters} misleads when processes of measurements \cite{JETP07Sh,Letter07} should be described. The magnetic field dependencies of the critical currents $I_{c+}(B)$ and $I_{c-}(B)$ were determined in \cite{JETP07Sh} by measuring periodically repeating current-voltage characteristics (a period of 10 Hz) in a slowly varying magnetic field $B_{sol}$ (a period of approximately 0.01 Hz) as follows. First, the condition that the structure was in the superconducting state was checked. Next, after the threshold voltage was exceeded (this voltage, set above induced voltages and noises of the measuring system, determined the minimum measurable critical current), magnetic field and critical current (with a delay of about $30 \ \mu s$) were switched on. This procedure allowed to measure sequentially critical currents in the positive $I_{c+}$  and negative $I_{c-}$  directions with respect to the external measuring current $I_{ext}$. Measurements of one $I_{c+}(B)$ or $I_{c-}(B)$  dependence (1000 values) took about 100 s. 

Thus, the ring (loop) was switched between superconducting and normal states at each $I_{c+}$ or $I_{c-}$ measurement by the measuring current $I_{ext}$ varying periodically between $I_{ext} < - I_{c-}(B)$ and $I_{ext} > +I_{c+}(B)$, see Fig.1. The measurements \cite{JETP07Sh} was made only in the temperature region $T < 0.991T _{c}$ where the current-voltage characteristics of single aluminium ring exhibit hysteresis and a sharp transition of the entire structure both to the normal and superconducting state, Fig.1. Because of the requirement that the complex pair wave function $\Psi  = |\Psi |exp(i\varphi )$ must be single-valued $\oint _{l}dl\nabla \varphi  = \oint _{l}dl (mv + qA)/\hbar = 2\pi n$ the pair momentum $mv$ 
$$\oint _{l}dl mv = 2\pi \hbar (n - \frac{\Phi }{\Phi _{0}})   \eqno{(1)}$$
(see the relation (1) in \cite{JETP07Sh} or (2) in \cite{Letter07}) and the persistent current $I_{p} = s2en _{s}v  = I_{p,A}2 (n - \Phi /\Phi _{0})$ circulating in the ring can not be equal zero in superconducting state when the magnetic flux $\Phi = BS + LI_{p} \approx BS \neq \Phi _{0}$ inside $l$. Here $\Phi _{0} = 2\pi \hbar /q$ is the flux quantum; $I _{p,A} = q \hbar/2mr\overline{(s n _{s})^{-1}} $ is the amplitude of the persistent current oscillations (when $n - \Phi /\Phi _{0}$ changes between -0.5 and 0.5) in a ring with section $s$ and pairs density $n _{s} = |\Psi |^{2}$ which may vary along the ring circumference $l$; $\overline{(s n _{s})^{-1}} = l^{-1}\oint _{l}dl (s n _{s})^{-1}$ \cite{PRB2001}. The quantum number $n$, describing the angular momentum of each $m _{p} =  \oint _{l}dl mv/2\pi = \hbar (n - \Phi /\Phi _{0})$ and all superconducting pairs $M _{p} = N _{s} m _{p} = (2m/q)I _{p}S$, is any integer number according to the requirement of quantization. But the measurements \cite{JETP07Sh,JETP07} indicate that the same quantum number $n$ is chosen almost always at each ring transition in superconducting state in a give magnetic field $\Phi \approx BS$. Different numbers, $n$ and $n+1$, or  $n$ and $n-1$ are chosen only in peculiar rings \cite{2statesLT34}. It is observed because of the predominate probability $P _{n} \propto \exp{- E_{n}/ k _{B}T} $ of the  permitted state (1) with minimal energy 
$$E_{n} = \oint _{l} dl sn _{s} \frac{mv _{n}^{2}}{2} =  I _{p,A}\Phi _{0} (n - \frac{\Phi }{\Phi _{0}})^{2}  \eqno{(2)}$$
The permitted state spectrum of real superconducting loop is strongly discrete $|E_{n+1} - E_{n}| = I _{p,A}\Phi _{0} \gg k _{B}T$: $ I _{p,A}\Phi _{0}/ k _{B} > 300 \ K$ at $T < 0.99T _{c}$ and typical value $I _{p,A} = 0.2 \ mA (1 -T/ T _{c})$ \cite{JETP07}. 

\begin{figure}
\includegraphics{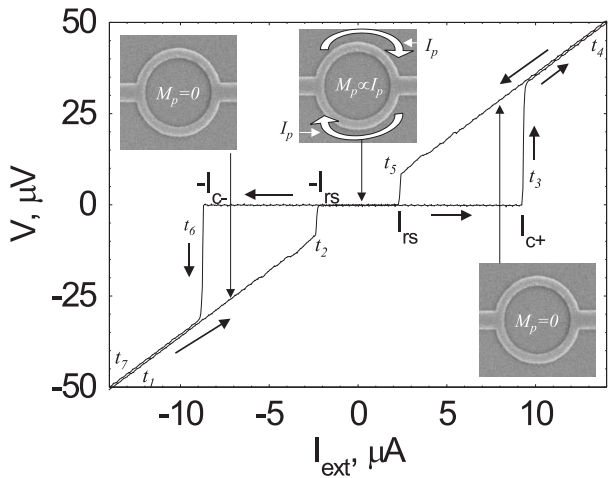}
\caption{\label{fig:epsart} A typical current-voltage characteristic (CVC) of single aluminium ring at $T < 0.99T _{c}$. The ring with radius $r \approx 2 \ \mu m$, the photos of which are shown, jumps from superconducting into normal state when the value of the external current $|I _{ext}|$ reaches the critical values $I _{c+}$ or $I _{c-}$ and comes back into superconducting state when the $|I _{ext}|$ value decries down to $I _{rs}$. The arrows along CVC indicate the direction of the $I _{ext}$ change in time. The ring photos are shown in order to accentuate that the angular momentum of the mobile charge carriers changes on a macroscopic value $|\Delta M _{p}| = (2m/q)|I _{p}|S$ at the transitions between superconducting and normal states because of the change of the circular electric current between $I _{cir} = I _{p}$ and  $I _{cir} = 0$.}
\end{figure}  

Thus, the permitted states with $|n - \Phi /\Phi _{0}| \leq 0.5$ is observed at the measurement of the critical current \cite{JETP07Sh}. These permitted states give predominate contribution to the values of the dc voltage and resistance at the observations of the dc voltage and the Little-Parks oscillation \cite{Letter07}. In contrast to our experiments \cite{JETP07Sh,Letter07} the authors \cite{Peeters} consider the permitted states with $|n - \Phi /\Phi _{0}| > 0.5$. In the case considered in \cite{Peeters} the vortex penetrates inside the loop or the quantum number $n$ changes when the critical current is reached in a loop segment with increasing applied current or magnetic field. Our experimental results \cite{JETP07} corroborate that the critical current of one-dimensional (i.e. $d, w \leq \xi (T)$, $dw \leq \lambda (T))$ loop is equal approximately the depairing current and therefore corresponds to the depairing velocity $v _{c} = \hbar /m \surd 3 \xi (T)$ \cite{Tink75}. The velocity (1) in a loop with approximately homogeneous section $s = dw$ reaches the critical value $v  = v _{c}$ at $|n - \Phi /\Phi _{0}| \approx l/2\pi \surd 3 \xi (T) $. Therefore the quantum number $n$ does not correspond, in general, to minimal kinetic energy (2), $|n - \Phi /\Phi _{0}| > 0.5$ and the hysteresis of magnetic dependence of loop magnetisation is observed \cite{nChMoler,nChGeim} at $ l/2\pi \surd 3 \xi (T) > 0.5$. The notion about vortex entry has a sense when the quantum number $n$ changes in \cite{nChMoler,nChGeim} on $\Delta n = 1, 2, 3,\cdot \cdot \cdot $ at $v  \approx v _{c}$, because measurable parameters connected with the pair momentum (1) change in this case. But it is absurd to say that $n = BS/\Phi _{0}$ vortices penetrate into a loop at its transition to superconducting state in the magnetic field  $B = n\Phi _{0}/S$. It is more correct to say that a qualitative transition from continuous to strong discrete spectrum of permitted states of the mobile charge carriers takes place in this case, see Fig1 in \cite{PRA2007QF}. The authors \cite{Peeters} could not have even a possibility to describe our experimental results \cite{PM2001,JETP07Sh,Letter07} using the time-dependent Ginzburg-Landau (GL) equation because the time-dependent GL theory can not describe this qualitative transition between states with different connectivity of wave function \cite{PRA2007QF}. 

\section{Macroscopic change of angular momentum because of the quantization}
A careful consideration of the process of the $I_{c+}, I_{c-}$ measurement \cite{JETP07Sh,JETP07} reveals a puzzle which is observed also in the Meissner effect. J.E. Hirsch wonders fairly \cite{Hirsch2010} that this puzzle is ignored: {\it "Strangely, the question of what is the 'force' propelling the mobile charge carriers and the ions in the superconductor to move in direction opposite to the electromagnetic force in the Meissner effect was essentially never raised nor answered"}. It is obvious that the velocity of the mobile charge carriers, the circular current $I _{cir}$ and the angular momentum $M _{p} = (2m/q)I _{cir}S$ change at the transition both in normal state at $|I_{ext}| > I_{c+}, I_{c-}$ and in superconducting state at  $|I_{ext}| < I_{rs}$, Fig.1. The angular momentum changes on a macroscopic value $|\Delta M _{p}| = (2m/q)|I _{p} - 0|S$ equal $|\Delta M _{p}| \approx 3 \ 10^{14} \hbar (1 -T/ T _{c})2|n - \Phi /\Phi _{0}|$ at the square $S = \pi r^{2} \approx 14 \ \mu m^{2}$ and the amplitude $I_{p,A} \approx 0.2 \ mA (1 -T/ T _{c})$ of the persistent current $I_{p}$ of the rings measured in \cite{JETP07Sh,JETP07}. It is well known that the circular current $I _{cir}(t) = I_{p}\exp -t/\tau_{RL}$ decays in a ring with an inductance $L$ a non-zero resistance $R > 0$ during the relaxation time $\tau_{RL} = L/R$ because of the dissipation force $mdv/dt =F _{dis}$ acting between electrons and the crystalline lattice of ions. Therefore there is not a problem with the angular momentum change from $M _{p} = (2m/q)I _{p}S$ to $M _{p} = 0$ after the transition in normal state. The current decays down to zero at the measurements in \cite{JETP07Sh,JETP07} because of the infinitesimality of the relaxation time $\tau_{RL} = L/R _{n} < 10^{-12} \ s$ (at $L \approx 10^{-12} \ H$ and $R _{n} > 15 \ \Omega $ of the rings) in comparison with the measurement time $t \approx 0.01 \ s$ of one current-voltage characteristic. 

All theories of superconductivity and quantum mechanics predict the change from $M _{p} = 0$ to $M _{p} = (2m/q)I _{p}S$ after the transition in superconducting state because  superconducting state with the pair angular momentum $m _{p} =  \oint _{l}dl mv/2\pi = 0$ is forbidden at $\Phi  \neq n\Phi _{0}$ (1). But no theory can say what is the 'force' propelling the mobile charge carriers in this case. No theory can say also how quickly the average velocity of the mobile charge carriers should change from $v = 0$ to $v = (\hbar /mr)(n - \Phi /\Phi _{0}) $ equal $\approx 14 \ m/s $ at the ring radius $r \approx 2 \ \mu m$ and $n - \Phi /\Phi _{0} \approx 1/2$. The basis of the theoretical investigation \cite{Peeters}, the time-dependent Ginzburg-Landau (GL) theory, includes the GL relaxation time $\tau _{GL}$ describing the change in time of the density of superconducting pairs $n _{s}$. But the 'force' propelling the mobile charge and a time of the velocity change \cite{PRA2007QF} can not be deduced from the time-dependent GL theory or any other theory of superconductivity known now. Therefore the authors \cite{Peeters} could consider theoretically only the case which differs in essence from the one of our measurements of the critical current \cite{JETP07Sh}. 

\section{The dc voltage and the Little-Parks oscillations}  
Their theoretical consideration \cite{Peeters} can not have also any relation to the observations of the dc voltage \cite{Physica1967,PM2001,Letter07} and the Little-Parks oscillations \cite{Letter07}. These observations reveal obvious paradox. It is well known that an electrical current must rapidly (during the relaxation time $\tau_{RL} = L/R$) decay in a loop with a resistance $R > 0$ if magnetic flux inside the ring does not change in time $d\Phi /dt = 0$. But the observations of the quantum oscillations of the resistance, $\Delta R \propto  \overline{I_{p}^{2}}$ \cite{LP1962,Letter07,toKulik2010} and magnetic susceptibility $\Delta \Phi _{Ip} = L\overline{I_{p}}$ \cite{PCScien07} give evidence that the persistent current can not decay in spite of non-zero resistance without the Faraday electrical field $-dA /dt = 0$. Moreover the measurements of the Little-Parks oscillations at low values of the measuring current $I _{ext}$ \cite{Letter07,toKulik2010} gives evidence that the persistent current can flow against electric field. The same paradox is obvious also in the observations of the dc voltage oscillations \cite{Physica1967,PM2001,Letter07,JETP07,Letter2003,PLA2012PV}. These paradoxes are direct consequence of the puzzle considered above. 

There is important to note that both the persistent current and the loop resistance are not zero on average in time $\overline{I_{p}} = \Theta ^{-1}\int _{0}^{\Theta }dt I _{p} \neq 0$, $\overline{R} = \Theta ^{-1}\int _{0}^{\Theta }dt R > 0$ in the process of the critical current measurements \cite{JETP07Sh,JETP07}. For case of the real measurements \cite{JETP07Sh} $\Theta $ is a time longer than the period $\approx 0.1 \ s$ of measuring current variation and much shorted than the period $\approx 100 \ s$ of magnetic field variation. According to the current-voltage characteristic shown on Fig.1 the ring resistance equals the resistance in normal state $R = R _{n}$ during the time from $t \approx t _{1}$ to $t \approx t _{2}$, from $t \approx t _{3}$ to $t \approx t _{5}$, from $t \approx t _{6}$ to $t \approx t _{7}$, and $R = 0$ during the time from $t \approx t _{2}$ to $t \approx t _{3}$ and from $t \approx t _{5}$ to $t \approx t _{6}$ when $I_{p} = I_{p,A}2 (n - \Phi /\Phi _{0})$. Therefore these values on average in time should be equal  $\overline{R} \approx R _{n}( t _{2} - t _{1} + t _{5} - t _{3} + t _{7}- t _{6})/( t _{7}- t _{1}) > 0$ and $\overline{I_{p}} \approx \overline{I_{p,A}}2 \overline{(n - \Phi /\Phi _{0})}( t _{3} - t _{2} + t _{6} - t _{5})/( t _{7}- t _{1}) \neq 0$ when $\overline{(n - \Phi /\Phi _{0})} \neq 0$. 

The average velocity of mobile charge carriers circulating in the loop changes from $\oint dl v = (2\pi \hbar /m) (n - \Phi /\Phi _{0})$ to $\oint dl v = 0$ because of the dissipation force $mdv/dt = F _{dis} $ at each transition in normal state at $|I _{ext}| \approx I _{c+}, I _{c-}$ and from $\oint dl v = 0$ to $\oint dl v = (2\pi \hbar /m) (n - \Phi /\Phi _{0})$ because of the quantization (1) at each transition in superconducting state at $|I _{ext}| \approx I _{rs}$, Fig.1. 

The latter explains why the persistent current can not decay in spite of non-zero dissipation \cite{PRA2007QF} for example in the Little-Parks effect $\Delta R(\Phi /\Phi _{0}) \propto \overline{I _{p}^{2}}$ \cite{Letter07,toKulik2010}. The dissipation power equal $\overline{RI _{cir}^{2}}$ without external current $I _{ext} = 0$  may be weak $\overline{RI _{cir}^{2}} = \Theta ^{-1}\int _{0}^{\Theta }dt RI _{cir}^{2}(t) = \Theta ^{-1}\int _{0}^{\Theta }dt RI_{p}^{2}\exp -2t/\tau_{RL} \approx L\overline{I_{p}^{2}}f _{sw}/2 \ll \overline{R}\overline{I_{p}^{2}}$ at a low frequency $f _{sw} \ll 1/\tau_{RL} = R _{n}/L$ of switching between superconducting and normal state. But it can not be zero at $\overline{R} > 0$ and $\overline{I_{p}}$ as well as the dissipation force $\oint dl \overline{F _{dis}} = \oint dl \Theta ^{-1}\int _{0}^{\Theta }dt F _{dis}dt = \Theta ^{-1}\int _{0}^{\Theta }dt d(\oint dl mv)/dt = \Theta ^{-1}\sum _{sw}[(\oint dl mv) _{t =  \Theta } - (\oint dl mv) _{t =  0}] \approx -2\pi \hbar  (\overline {n} - \Phi /\Phi _{0}) \neq 0$ when $\overline {n} - \Phi /\Phi _{0} \neq 0$. The obvious equality $2\pi \hbar (n - \Phi /\Phi _{0}) - 2\pi \hbar (n - \Phi /\Phi _{0}) = 0$ taken on average in time $\Theta $ gives a "force" balance 
$$ \frac{2\pi \hbar }{m}(\overline{n} - \frac{\Phi }{\Phi _{0}}) f _{sw} + \oint dl \overline{F _{dis}} = lF _{q} + \oint dl \overline{F _{dis}} = 0  \eqno{(3)}$$
where $(2\pi \hbar /m) (n - \Phi /\Phi _{0}) f _{sw}/l = F _{q}$ is the momentum change in a time unity of mobile charge carrier because of the quantization called in \cite{PRB2001} "quantum force". The quantum force replaces formally the force $-qdA/dt$ of the Faraday electric field $-dA/dt$ and can explain \cite{PRA2007QF} the paradoxical phenomena observed at the measurements of the Little-Parks \cite{Letter07,toKulik2010} and the dc voltage \cite{Physica1967,PM2001,Letter07,JETP07,Letter2003,PLA2012PV} oscillations. The quantum force can not be deduced on the base of the time-dependent Ginzburg-Landau theory. Therefore the behaviour of the output voltage on the applied field observed in \cite{Physica1967,PM2001,Letter07}, in particular shown on  Fig. 4 in \cite{Letter07} can not be similar with anything obtained in \cite{Peeters}. 

We regret that the authors \cite{Peeters} could not understand totally the essence of the behaviour of the output voltage shown on Fig. 4 in \cite{Letter07}. It is clearly written in \cite{Letter07} that because of the presence of alternating-sign $V_{p}(\Phi /\Phi _{0})$ oscillations (shown on Fig. 5), the total output  voltage (shown on Fig. 4) is the sum $V(\Phi /\Phi _{0}) = R_{0}I_{ext} + \Delta R(\Phi /\Phi _{0})I_{ext} + V_{p}(\Phi /\Phi _{0})$ of the voltage $R_{0}I_{ext} + \Delta R(\Phi /\Phi _{0})I_{ext} $ depending on the external dc current $I_{ext}$ and $V_{p}(\Phi /\Phi _{0})$.  The Little-Parks oscillations $\Delta R(\Phi /\Phi _{0})I_{ext} \propto \overline{I _{p}^{2}} \propto \overline{(n - \Phi /\Phi _{0})^{2}}$ have extremums at $\Phi = n\Phi _{0}$ and $\Phi = (n+0.5)\Phi _{0}$, see Fig.3 in \cite{Letter07} whereas the extremums of the alternating-sign oscillations $V_{p}(\Phi /\Phi _{0})$ are observed between $\Phi = n\Phi _{0}$ and $\Phi = (n+0.5)\Phi _{0}$, see Fig.5 in \cite{Letter07}. The amplitude of the Little-Parks oscillations $\Delta R(\Phi /\Phi _{0})$ shown on Fig.4 in \cite{Letter07} equals approximately $40 \ \Omega $ and of the $ V_{p}(\Phi /\Phi _{0})$ oscillations shown on Fig.5 in \cite{Letter07} equals approximately $600 \ nV$. Therefore the extremums of the total output voltage $V(\Phi /\Phi _{0})$ are observed at $\Phi = n\Phi _{0}$ and $\Phi = (n+0.5)\Phi _{0}$ when $|I_{ext}| \gg 600 \ nV/40 \ \Omega = 15 \ nA$, Fig.3 in \cite{Letter07}, and between $\Phi = n\Phi _{0}$ and $\Phi = (n+0.5)\Phi _{0}$ when $|I_{ext}| \ll 15 \ nA$, Fig.5 in \cite{Letter07}. No result of \cite{Peeters} can have any relation to the both cases.

\section{Conclusion} 
The publication \cite{Peeters} reveals the lack of understanding that existing theories of superconductivity can not provide a complete description of all quantum phenomena observed in superconductors. The force-free momentum change of the mobile charge carriers at the transition into superconducting state is not only puzzle which no theory of superconductivity can solve. We would like to draw readers attention on a paradoxical contradiction between theoretical prediction and results of measurements of magnetic dependencies of the critical current of asymmetric superconducting rings revealed in our works \cite{JETP07Sh,JETP07}. A simple theoretical consideration \cite{JETP07} based on the condition of quantization (1) predicts that the critical current of symmetric (i.e. with the same width $w _{w} = w _{n}$ and length $l _{w} = l _{n} = l/2$ of the ring arms) ring should oscillate in magnetic field $\Phi \approx BS$ in accordance with the relation $I _{c} = I _{c0} - 2 |I _{p}| = I _{c0} - 2I _{p,A}2|n - \Phi /\Phi _{0}|$. The pair velocity (1) and the persistent current corresponding to the minimal energy (2) should jump at $\Phi = (n+0.5)\Phi _{0}$ from $I _{p} = I _{p,A}2(-0.5) = - I _{p,A}$ to $I _{p} = I _{p,A}2(+0.5) = + I _{p,A}$ with the quantum number change from $n$ to $n+1$. This jump should not be observed at measurement of the critical current of symmetric ring because $|-I _{p,A}| =  |+I _{p,A}|$. But the jump of the critical current $\Delta I _{c} = I _{p,A}(w _{w}/w _{n} - w _{n}/w _{w})$ should be observed at measurement of the asymmetric ring with different width $w _{w} > w _{n}$, see Fig.19 in \cite{JETP07}, or different length $l _{w} > l _{n}$ \cite{NANO2011} of the ring arms. Our measurements \cite{JETP07Sh} as well as measurements of other authors of symmetric rings have corroborated the theoretical prediction. But measurements of asymmetric rings made first in our works \cite{JETP07Sh,JETP07} have revealed qualitative discrepancies between theoretical and experimental magnetic dependencies of the critical current. First of all we have discovered that the jump predicted because of the quantum number $n$ change is absent on the experimental dependencies measured on rings both with $w _{w} > w _{n}$ \cite{JETP07} and $l _{w} > l _{n}$ \cite{NANO2011}. These results are very strange because the observed periodicity can be connected only with the $n$ change and the quantum number can not change on a value lesser than unity according to the basic principles of quantum mechanics. Other paradoxical result discovered at measurement of rings with $w _{w} > w _{n}$ is the shift of the critical current oscillations $I_{c+}(\Phi /\Phi _{0}), I_{c-}(\Phi /\Phi _{0})$ with the appearance of ring asymmetry up to $\pm \Phi _{0}/4$ at  $w _{w}/w _{n} \geq 1.25$ \cite{JETP07Sh}. This shift of $I_{c+}(\Phi /\Phi _{0})$ and $I_{c-}(\Phi /\Phi _{0})$ in opposite direction provides the anisotropy $ I_{c,an} = I_{c+}(\Phi /\Phi _{0}+1/4) - I_{c-}(\Phi /\Phi _{0} - 1/4) \neq 0$ of the critical current of asymmetric ring and the explanation of the rectification effect \cite{JETP07}. But the $I_{c+}(\Phi /\Phi _{0}), I_{c-}(\Phi /\Phi _{0})$ extremums should be observed at $\Phi = n\Phi _{0}$ and $\Phi = (n+0.5)\Phi _{0}$ according to the condition of the quantization (1) and can not be observed at $\Phi = (n+0.25)\Phi _{0}$ and $\Phi = (n+0.75)\Phi _{0}$. The qualitative discrepancies between theoretical predictions and experimental results should be explained. The attempt by the authors \cite{Peeters} to explain at least the shift is completely unfounded.

\end{document}